\documentclass[11pt]{article}
\usepackage{graphicx}
\usepackage{amssymb}

\title{Interdigitation between surface-anchored polymer chains and an elastomer : consequences for adhesion promotion}
\author{T. Vilmin(1), C. Tardivat(1), L. L\'eger(1), H. Brown(2) and E. Rapha\"el(1) \\ \\
(1) Laboratoire de Physique de la Mati\`ere Condens\'ee\\
UMR 7125 CNRS and FR 2438 "Mati\`ere et Syst\`emes Complexes"\\
Coll\`ege de France\\
 11, Place Marcelin Berthelot, 75231 Paris Cedex 05, France\\ \\
 (2) Steel Institute, University of Wollongong, Northfields Avenue\\
 Wollongong N.S.W 2522, Australia
}

\begin{document}
\maketitle

\begin{abstract}
We study the adhesion between a cross-linked elastomer and a flat solid surface where polymer chains have been end-grafted. To understand the adhesive feature of such a system, one has to study both the origin of the grafted layer interdigitation with the network, and the end-grafted chains extraction out of the elastomer when it comes unstuck from the solid surface. We shall tackle here the first aspect for which we develop a partial interdigitation model that lets us analytically predict a critical surface grafting density $\sigma^{*} \simeq P^{\frac{1}{10}}N^{-\frac{3}{5}}$ beyond which the layer no longer interdigitates with the elastomer. We then relate this result with recent adhesion measurements.

\end{abstract}

\section{Introduction}

Surface-anchored polymer layers play an important role in adhesion \cite{Jones}\cite{Wool}. The key parameter is the degree of interdigitation between the surface layer and the bulk polymer system \cite{LRH99}\cite{Brown02}\cite{Brown96}.
Some years ago, de Gennes proposed an interesting analogy between the behavior of grafted chain penetrating an elastomer with $P$ monomers between cross-links, and the behavior of the same grafted chains immersed in a melt of $P$ monomers chains \cite{PGG94}. In the later situation, the melt chains screen out the excluded volume interaction by a factor $1/P$ \cite{DoiEdw}. In the former case, a similar screening occurs through the elastic deformation of the network, which has an elastic modulus $E \simeq kT/a^{3}P$, where $a$ is the typical size of a monomer. Let us figure out the simple case where each grafted chain ($N$ monomers per chain) is extended over a typical distance $L$. If $\sigma$ is the dimensionless surface grafting density($\sigma = a^{2}\Sigma$), the average volume fraction occupied by the grafted layer is $\phi_{av} \simeq \sigma Na/L$. In the melt case, when $\phi_{av}$ is small compared to unity, the Flory expression for osmotic free energy  per grafted chain can be aproximated by $(La^{2}/\sigma)\left(kT(1-\phi_{av})ln[1-\phi_{av}]/a^{3}P\right) \simeq kT(\sigma Na/L-1)(N/P)$. In the elastomer case, de Gennes showed that the swelling free energy per volume unit is $E\phi^{2}$, which gives the expression $(La^{2}/\sigma)\left(kT\phi_{av}^{2}/a^{3}P\right) \simeq kT(\sigma N^{2}a/LP)$ per grafted chain. As we can see, those two expressions are identical except for a constant. So, since the two other components of the grafted chain free energy are independent of the surrounding environment (see eq.~\ref{fanalog}, where the first term represents the elongation energy and the second one is due to the mean volume fraction gradient $\phi_{av} /L$, all numerical factor being ignored), the N-chain conformation is the same in both situation.

\begin{equation}
\frac{F}{kT} \simeq
\frac{L^{2}}{a^{2}N}+\frac{a^{2}N}{L^{2}}+\frac{\sigma N^{2}a}{PL}
\label{fanalog}
\end{equation}

The various regimes of interdigitation between a brush and an elastomer can be sketched by the diagram used for the brush interdigitation with a melt \cite{LRH99}. 
However, even if the network swelling free energy seems to favor the penetration of the layer, for $L\phi_{av}^{2}$ is a decreasing function of $L$, it is always a positive energy which will limit penetration if $\phi_{av}$ is too high. This is not true in the melt case where the Flory-Huggins free energy is negative when one takes into account the additive constant, and where interdigitation is predicted until $\phi_{av}$ reaches $1$ for $\sigma=1/P^{\frac{1}{2}}$. 
For this reason, Brochard-Wyart \textit{et al.} \cite{Broch94} (hereafter referred-to as BW) recognized that the analogy would break down before that $\sigma$ reaches $1/P^{\frac{1}{2}}$. They evaluated the limit grafting density beyond which penetration can not be total anymore around $\sigma=P/N^{\frac{3}{2}}$, where the network swelling free energy per grafted chain is equal to $kT$.
From this point they proposed a partial penetration regime, which goes from $\sigma=P/N^{\frac{3}{2}}$ to $\sigma=1/P^{\frac{1}{2}}$,  where each connector inserts only $n$ $(<N)$ monomers into the network. Their model is based on the assumption that $n$ adjusts itself in order to fix the swelling free energy per grafted chain at $kT$, and not on a study of the equilibrium states resulting from the minimization of the total free energy of the system.
We propose in the following paper to reexamine those partial penetration problems.

\section{Dry brush in contact with an elastomer}

\begin{figure}
\begin{center}
\includegraphics[width=6.5cm]{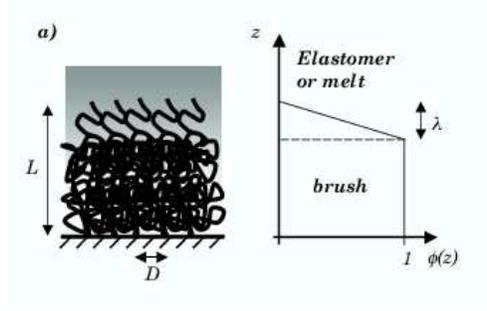}
\caption{Dry brush regime}
\label{f.1a}
\end{center}
\end{figure}

In order to consider those questions it is simpler to begin with the limit where the elastomer is in contact with a very dense grafted layer with almost no interdigitation, as Leibler and co-authors \cite{OUMS93} did for what they called a "dry brush" in contact with a melt.
In a dry brush regime the grafted layer interdigitates with the melt, or the network, only over a width $\lambda$, small compared to the thickness of the layer. This thickness is thus close to the minimum thickness $h_{0}=\sigma Na$ (see fig.~\ref{f.1a}). The free energy per grafted chain is still composed of three different terms. One can show \cite{OUMS93} that the elongation term is given by $F_{el}/kT \simeq \lambda^{2}/a^{2}N$ (where the constant term $h_{0}^{2}/a^{2}N$ has been subtracted) \footnote{This result can be simply derived by assuming that half of the chains are extended over the length $h_{0}-\lambda/2$, and the other half over the length $h_{0}+\lambda/2$. Note that assuming that all the grafted chains are equally stretched would lead to the stretching energy $F_{el}/kT \simeq \lambda h_{0}/a^{2}N$ (where again the term $h_{0}^{2}/a^{2}N$ has been subtracted).}. The gradient term, which can be renamed as the interfacial term, is given by $F_{int}/kT \simeq a/\sigma \lambda$ \cite{OUMS93}. Now, in the melt case, the Flory-Huggins expression for osmotic free energy is dominated by the lower order term : $F_{osm}/kT \simeq -\lambda/\sigma aP$. In the elastomer case, the de Gennes expression for swelling energy is : $F_{swell}/kT \simeq \lambda/\sigma aP$. Therefore, as suspected from the beginning, the analogy breaks down at high $\sigma$, in dry brush regimes. Actually, the osmotic pressure still favors interdigitation, whereas the network swelling energy now goes against interdigitation, for $\phi_{av}$ does not reduce when $\lambda$ increases, and this leaves the interfacial energy as the only term that drives interdigitation. 

\begin{equation}
\frac{F_{melt}}{kT} \simeq \frac{\lambda^{2}}{a^{2}N} +
\frac{a^{2}}{\sigma}\frac{1}{a\lambda} - \frac{a^{2}}{\sigma}\frac{\lambda}{a^{3}P}
\label{fdrymelt}
\end{equation}

Leibler and co-authors showed two distinct dry brush regimes in the melt case (from eq.~\ref{fdrymelt}): if $\sigma>N/P^{\frac{3}{2}}$ the osmotic energy is weak compared with the interfacial energy, and $\lambda \simeq a(N/P)^{\frac{1}{3}}$. If $\sigma<N/P^{\frac{3}{2}}$ the osmotic energy is large compared with the interfacial energy, and $\lambda \simeq aN/\sigma P$. A remarkable fact is the perfect continuity of $L$, $\lambda$ and $\phi_{av}$ through the transition $\sigma=1/P^{\frac{1}{2}}$ between total interdigitation regimes and dry brush regimes for a melt.

\begin{equation}
\frac{F_{elasto}}{kT} \simeq \frac{\lambda^{2}}{a^{2}N} +
\frac{a^{2}}{\sigma}\frac{1}{a\lambda} + \frac{a^{2}}{\sigma}\frac{\lambda}{a^{3}P}
\label{fdryelasto}
\end{equation}

When the dry brush is in contact with an elastomer, two regimes also exist (from eq.~\ref{fdryelasto}): if $\sigma>N/P^{\frac{3}{2}}$ the network swelling energy is weak compared with the grafted chains' elastic energy, and $\lambda \simeq a(N/P)^{\frac{1}{3}}$, which is the same interface width as in the melt case. If $\sigma<N/P^{\frac{3}{2}}$ the network swelling energy is large compared with the chains' elastic energy, and $\lambda \simeq aP^{\frac{1}{2}} = \lambda_{0}$ which is the elastomer network mesh size. What is now to be noticed is that there is no possible continuous transition of $L$, $\lambda$ and $\phi_{av}$ between this dry brush regime and the total interdigitation regimes. So, something is still to be understood about this transition, and we are back to the partial interdigitation assumption.

\begin{figure}
\begin{center}
\includegraphics[width=6.5cm]{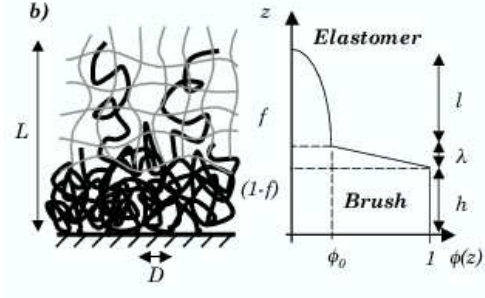}
\caption{Partial interdigitation state.}
\label{f.1b}
\end{center}
\end{figure}

\section{Partial interdigitation}

BW's model is the simplest partial penetration model, but if one writes the total free energy by grafted chain as a function of $n$, one can show that the interface between the elastomer and the dense part of the brush composed by the $(N-n)$ monomers by chain that do not penetrate has an important energetic cost that leads to an equilibrium state with $n=0$, that is to say a dry brush, whatever the value of $\sigma$. Thus this model does not help to understand the total interdigitation-dry brush transition.

The second simplest partial penetration model one can build consists in taking a fraction $f$ of the grafted chains that almost fully penetrates the network, and another $(1-f)$ fraction that is sandwiched in between the solid surface and the network with which it interdigitates over the length $\lambda$ (see fig.~\ref{f.1b})\,\footnote{BW also considered this situation, but did not retain it as their simplified calculus of the free energy led to total penetration until $\sigma$ reaches $1/P^{\frac{1}{2}}$.  Our study, carried out a little bit farther, gave us a different result.}. We can determine $\lambda$ self consistently, as $\lambda$, $\phi_{0}$ (which is the average volume fraction occupied by the $f$-chains) and $f$ are the three parameters the free energy per grafted chain depends on. As for the brush that fully penetrates or a dry brush, the free energy is made up a stretching term, a confinement term and a swelling term. The stretching energy is the average of the $f$-chains elastic energy and the $(1-f)$ stretching energy:

\begin{equation}
\frac{F_{el}}{kT} \simeq
f\frac{L^{2}}{a^{2}N} + (1-f)\frac{(h+\lambda)^{2}}{a^{2}N} -
\frac{h_{0}^{2}}{a^{2}N} 
\end{equation}

Here $L$ and $(h+\lambda)$ can be replaced using the volume conservation :  $L \simeq fh_{0}/\phi_{0}$ and $(h+\lambda) \simeq (1-f)h_{0}/(1-\phi_{0})$. The interfacial energy is also the average between a $f$-chains' term and a $(1-f)$-chains' term :

\begin{equation}
\frac{F_{int}}{kT} \simeq \frac{a^{2}}{\sigma}\left( f \frac{\phi_{0}}{al} + (1-f)\frac{(1-\phi_{0})}{a\lambda}\right)
\end{equation}

Then the network swelling energy is given by a de Gennes' expression integration over the thickness of the brush :

\begin{equation}
\frac{F_{swell}}{kT} \simeq
\frac{a^{2}}{\sigma}\left(\frac{\lambda}{a^{3}P} + \frac{l\phi_{0}^{2}}{a^{3}P} \right)
\end{equation}

In the following study we will consider that $P^{\frac{1}{2}}/N \ll \sigma \ll N/P^{\frac{3}{2}}$, so that       $\lambda \sim \lambda_{0} \ll h_{0}$, and that we can neglect the $(1-f)$-chains' stretching energy. Provided that $f>\phi_{0}$, volume conservation leads to $l \simeq h_{0}(f-\phi_{0})/(\phi_{0}(1-\phi_{0}))$. Given those relations, we can calculate the equilibrium value of $\lambda$ :

\begin{equation}
\lambda \simeq
\lambda_{0}(1-\phi_{0})^{\frac{1}{2}}(1-f)^{\frac{1}{2}}
\end{equation}

and we obtain a two parameters free energy per grafted chain :

\begin{equation}
\frac{F(f, \phi_{0})}{kT} \simeq 2\frac{(1-\phi_{0}
)^{\frac{1}{2}}(1-f)^{\frac{1}{2}}}{\sigma P^{\frac{1}{2}}} +
\sigma^{2}N\frac{f^{3}}{\phi_{0}^{2}} +
\frac{f\phi_{0}^{2}(1-\phi_{0})}{\sigma^{2}N(f-\phi_{0})} +
\frac{N\phi_{0}(f-\phi_{0})}{P(1-\phi_{0})}
\end{equation}

\begin{figure}
\begin{center}
\includegraphics[width=10cm]{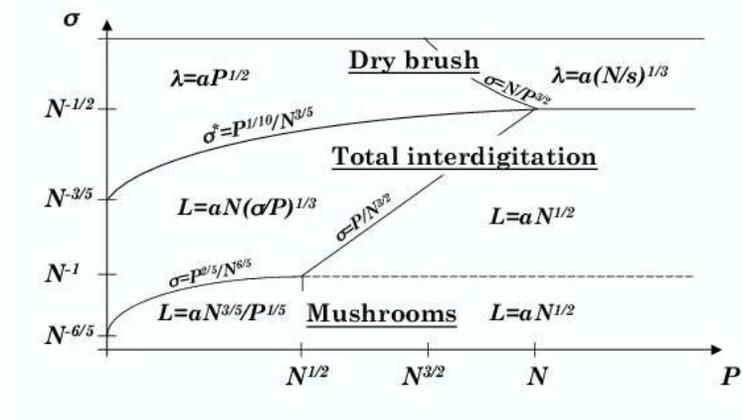}
\caption{Interdigitation regimes of a grafted layer and an elastomer corresponding to various $(\sigma, P)$ couples ($\sigma$ is the surface grafting density adimentionalized with $a$, $P$ is the number of monomers between cross-links, and number of monomers per grated chain $N$ is fixed). The dashed line in the left hand side dry brush regime is the limit grafting density for interdigitation in the melt case.}
\label{f.2}
\end{center}
\end{figure}

The first term is the $(1-f)$-chains' interfacial energy, and the three other terms are the $f$-chains' energy. The two local minimums of this $(f, \phi_{0})$ function can be determined analytically. The first one correspond to $\phi_{0}=\sigma^{\frac{2}{3}}P^{\frac{1}{3}}$ and $f=1$, which is the classical total interdigitation state, with $F_{tot}\simeq(\sigma/P)^{\frac{2}{3}}NkT$ \cite{LRH99}.
The second local minimum correspond to $\phi_{0}\simeq f(1-P^{\frac{1}{2}}/\sigma N)$, leading to  $l \simeq \lambda_{0}(1+f)/2 \leq \lambda_{0}$. Then the $f$-chains have a similar non-stretched conformation as the $(1-f)$-chains, and the interpenetration is weak. When $f$ varies from $0$ to $1$ the free energy of this local minimum varies from $0.75F_{dry}$ to $F_{dry}$ ($F_{dry}\simeq kT/ \sigma P^{\frac{1}{2}}$), the minimum corresponding to $f=1/2$; this is thus a dry brush, just  slightly more refined than the dry brush model we developed earlier.
Those two local minimums, separated by a gap, are the only two possible equilibrium states; whatever the value of $\sigma$ is, partial interdigitation is not the lowest free energy state. When $\sigma < \sigma^{*} \simeq P^{\frac{1}{10}}/N^{\frac{3}{5}}$, the lowest free energy state is total interdigitation. Then, if $\sigma > \sigma^{*}$, the lowest free energy state is the dry brush state. This result, combined with the previously set up idea that BW's model can't give partial penetration, leads us to assume that partial penetration never appears as a stable state\,\footnote{Partial interdigitation does not happen because of the coupling (induced by volume conservation) between the thickness $h$ (see fig.~\ref{f.2}) and the amount of penetrating grafted chains. Note that at the crack tip, this coupling plays a less important role and it can then be relevant to study partial penetration assuming $h$ to be constant (with a value of order $h_{0}$ independent of the amount of penetrating grafted chains, and an interface width of order $\lambda_{0}$). Such a simplified approach would give $\frac{F}{kT} \simeq \frac{a^{2}m}{\lambda_{0}h} + \frac{h^{2}}{a^{2}m} + n\left(\frac{\sigma}{P}\right)^\frac{2}{3}$ as a free energy of a chain penetrating on $n$ monomers, with $m=N-n$ monomers out. Minimization gives $m=\frac{h}{a}(\frac{a^{2}}{\lambda_{0}h}-\left(\frac{\sigma}{P}\right)^{\frac{2}{3}})^{-\frac{1}{2}}$, which is finite for $\sigma<\sigma^{*}$. Thus, grafted chains penetrate partially at the crack tip. This shows how total interdigitation can propagate from a crack tip below $\sigma^{*}$.}.

The new interdigitation limit $\sigma^{*} \simeq P^{\frac{1}{10}}/N^{\frac{3}{5}}$ is drawn on the graph (see fig.~\ref{f.2}). Unlike the limit interdigitation grafting density $\sigma \simeq 1/P^{\frac{1}{2}}$ in the melt case, $\sigma^{*}$ is an increasing function of $P$, which is very intuitive, for the bigger $P$, the softer the elastomer, and the easier the grafted chains can penetrate the network until they reach high volume fraction. Note that this limit has not the same meaning as the other limits of the graph, as the transition between $L$ and $\lambda$ is discontinuous. It can be seen as a predominance limit between a regime where total interdigitation is the most stable state, and a regime where a dry brush is more stable; thermal fluctuation always allowing the grafted layer to switch from one state to another, passing through partial penetration. Irregularities of connectors surface density may play an important role about this.

\section{Adhesion measurments}
\begin{figure}
\begin{center}
\includegraphics[width=6.0cm]{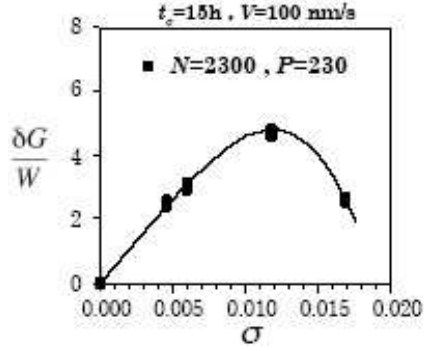}
\caption{Adhesion enhancement as a function of surface grafting density for a PDMS  layer of $N=2300$ chains and a PDMS elastomer of $P=230$ cross-links index. The full line is only a guide for the eyes.}
\label{f.3a}
\end{center}
\end{figure}

In order to test how interdigitation between a brush and an elastomer can enhance adhesion at such interfaces, we have conducted series of experiments on model systems based on polydimethylsiloxane (PDMS) grafted surface put into contact with PDMS crosslinked elastomers . Grafted PDMS layers were formed on the surface of silicon wafers with a SiH carpet on top. In order to fully protect the silica surface from direct interactions between monomers of the elastomer and the silanol sites of the  surface, bimodal brushes were grafted. A mixture of two different molecular weights of monovinyl terminated PDMS, one smaller than the critical molecular weight between entanglements and one larger, is put into contact with the SiH modified surface. By varying the relative concentration of long and short chains in the mixture, one could control the thickness of the grafted layer, and thus the surface density of long chains in the grafted layer.Model crosslinked elastomers were synthesized using the same hydrosililation reaction as the grafted brushes. The crosslinking reaction was optimized and the stoechiometric ratio of SiH over Si-vinyl concentration chosen in order to minimize the number of dangling chains \cite{Am03}. 
A home made JKR apparatus \cite{Deru98}\cite{Am03} was used to form a contact between the grafted layers and microlenses of optimized elastomers, and then to characterise the fracture toughness at very low advancing fracture velocity (in the range $1nm/s$ to $10\mu m/s$), as a function of both contact time and surface grafting density of long chains in the grafted layer and molecular weight between crosslinks in the elastomer. First, on dense grafted layer of short chains, no hysteresis was observed between loading and unloading in the JKR apparatus, even after a contact time of several days. The measured adhesion energy $W$ was just twice the surface tension of PDMS, as expected for the contact between two PDMS surfaces with no significant interdigitation across the interface. On the contrary, as soon as long chains were present at the interface, adhesion hysteresis was observed, and which increase with contact time, first rapidly in the first hours of contact and then very slowly, during days and month. Typical data, obtained for one polymerization index between crosslinks, P = 230, one polymerisation index of the long grafted chains, N = 2300, (short grafted chains with Ns = 68), after a contact time of 15hours, and for different grafting densities are reported in figure \ref{f.3a}, in terms of adhesion enhancement $\delta G(\sigma) = G(\sigma) - W$ normalized by the thermodynamic work of adhesion $W$ measured on the short grafted layer. We do see that the long chains are able to enhance adhesion significantly. It is also clear that an optimum surface density exists, and that one looses adhesion enhancement for large enough $\sigma$.

\begin{figure}
\begin{center}
\includegraphics[width=6.0cm]{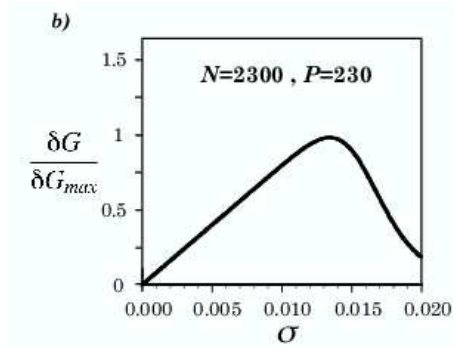}
\caption{Theoretical prediction for adhesion enhancement ($N=2300$ and $P=230$).}
\label{f.3b}
\end{center}
\end{figure}

A naive idea would be to say that the above adhesion enhancement comes from the need to reduce the degree of interdigitation at the crack tip in order to allow the elastomer to come unstuck from the brush. This would only add to the thermodynamic work of adhesion $kT/a^{2}$, which is of order $W$ and independent of $\sigma$.
Therefore, in order to fully analyze these adhesion measurements, one would need a detailed description of the dissipative pull-out mechanism taking place at the fracture tip. However, since experimentally $\delta G/W$ is a linear $\sigma$ function at low $\sigma$'s, we can assume that $\delta G/W$ is of the form $\delta G/W = \alpha \sigma$ whenever the grafted layer is in a total interdigitation state (for a discussion of the form of $\alpha$, see \cite{RDG92}). Experimental $\delta G/W(\sigma)$ data plots show a decrease at high $\sigma$, so we also assumed that $\delta G/W$ is approximately nil when the layer is in a dry brush state (only the fully penetrating chains participate in the adhesion enhancement). Then, we took into account thermal fluctuations, writing $\delta G/W(\sigma)$ as the thermodynamic average between the surface adhesion of a total interdigitation layer and a dry brush (see eq.~\ref{deltaG}).

\begin{equation}
\frac{\delta G}{W} =
\alpha\sigma\frac{\exp{\left[-\frac{F_{tot}}{kT}\right]}}{\exp{\left[-\frac{F_{tot}}{kT}\right]}+\exp{\left[-\frac{F_{dry}}{kT}\right]}}
=\alpha\sigma\frac{\exp{\left[\left(1-\left(\frac{\sigma}{\sigma^{*}}\right)^{\frac{5}{3}}\right)\frac{1}{\sigma
P^{\frac{1}{2}}}\right]}}{1+\exp{\left[\left(1-\left(\frac{\sigma}{\sigma^{*}}\right)^{\frac{5}{3}}\right)\frac{1}{\sigma
P^{\frac{1}{2}}}\right]}}
\label{deltaG}
\end{equation}

\begin{figure}
\begin{center}
\includegraphics[width=11cm]{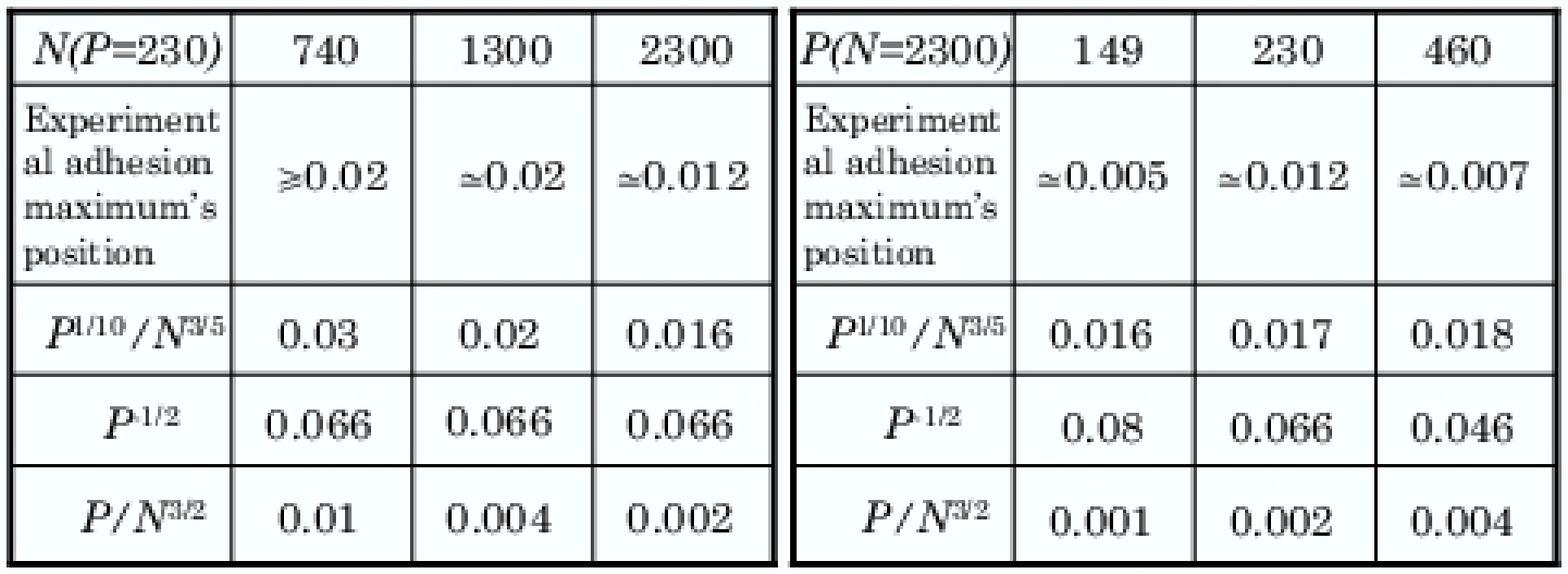}
\caption{Comparison of experimental maximum adhesion's grafting density with the present theoretical prediction and former theoretical predictions.}
\label{f.4}
\end{center}
\end{figure}

Figure \ref{f.3b} represents $\delta G(\sigma)$ in the case $N=2300$ and $P=230$. The maximum surface adhesion energy is reached for $\sigma \simeq \sigma^{*}$, and then $\delta G$ decreases over the characteristic grafting density range $\sigma^{*}(P/N)^{\frac{3}{5}}$. Thermal fluctuations are thus enough to smoothen  the total interdigitation-dry brush transition. One can see, comparing with experimental data, that the agreement is fairly good. In figure \ref{f.4} we compare case by case $\sigma^{*}$ with the position of experimental adhesion maximum. One can see that $\sigma^{*}$ is in the correct range, which is not the case for $1/P^{\frac{1}{2}}$ or $P/N^{\frac{3}{2}}$. Nevertheless, more work is required to fully confirm $\sigma^{*}$'s scaling low on $N$ and $P$.

To conclude, we have shown that the limit surface grafting density, beyond which almost no interdigitation occurs, can be analytically calculated. This limiting connector surface density, $\sigma^{*} \simeq P^{\frac{1}{10}}/N^{\frac{3}{5}}$, apparently also corresponds to the grafting density for the maximum adhesion. Although the knowledge of the several interdigitation regimes lets one understand the general feature of elastomer-grafted layer adhesion, a clear description of connectors extraction microscopic process is necessary if one pretends to forecast the value of the maximum surface adhesion energy that can be reached.



\begin{thebibliography}{0}

\bibitem{Jones}
R. A. L. Jones, R. W. Richards, Polymers at Surfaces and Interfaces,
Cambridge University Press (1999).

\bibitem{Wool}
R. P. Wool, Structure and Strength of Polymer Interfaces, Hanser/Gardner Publications, NY (1995).

\bibitem{LRH99}
L. L\'eger ,  E. Raphael ,  H. Hervet, Advances in Poly. Sci.138, 185 (1999).
  
\bibitem{Brown02}
C.  Creton ,  E. J. Kramer ,  H. R. Brown ,  C. Y. Hui
, Adv. Polym. Sci. 156, 53, (2002).

\bibitem{Brown96}
H. R. Brown
, Physics World 9, 38, (1996).

\bibitem{PGG94}
P. G. de Gennes
, C. R. Acad. Sci. 318 II, 165, (1994).

\bibitem{DoiEdw}
M. Doi ,  S. F. Edwards, The Theory of Polymer Dynamics, Oxford Science Publications (1986).

\bibitem{Broch94}
F. Brochard-Wyart ,  P. G. de Gennes ,  L. L\'eger ,  Y. Marciano ,  E. Raphael,
J. Phys. Chem., 98, 9405 (1994). See also the related work of C. Ligoure
, Macromol. 29, 5459 (1996).

\bibitem{OUMS93}
L. Leibler ,  A. Ajdari ,  A. Mourran ,  G. Coulon ,  D. Chatenay
, in 1993 OUMS conf. on Ordering in Macromolecular Systems, Osaka, OUMS, Springer Verlag, Berlin (1994).

\bibitem{Am03}
N. Amouroux ,  L. L\'eger
, Langmuir 19, 1396 (2003)

\bibitem{Deru98}
M. Deruelle ,  H. Hervet ,  G. Jandeau ,  L. L\'eger
, J. Adhesion Sci. Technol., 12, 225, 1998

\bibitem{RDG92}
E. Raphael ,  P. G. de Gennes
, J. Phys. Chem. 96, 4002 (1992).

\end{thebibliography}
 \end{document}